\documentstyle[12pt,epsfig]{article}

\newlength{\dinwidth}
\newlength{\dinmargin}
\begin{document}
\def\bold#1{\setbox0=\hbox{$#1$}%
     \kern-.025em\copy0\kern-\wd0
     \kern.05em\%\baselineskip=18ptemptcopy0\kern-\wd0
     \kern-.025em\raise.0433em\box0 }
\def\slash#1{\setbox0=\hbox{$#1$}#1\hskip-\wd0\dimen0=5pt\advance
         to\wd0{\hss\sl/\/\hss}}
\newcommand{\be}{\begin{equation}}
\newcommand{\ee}{\end{equation}}
\newcommand{\bea}{\begin{eqnarray}}
\newcommand{\eea}{\end{eqnarray}}
\newcommand{\nn}{\nonumber}
\newcommand{\dd}{\displaystyle}
\newcommand{\bra}[1]{\left\langle #1 \right|}
\newcommand{\ket}[1]{\left| #1 \right\rangle}
\newcommand{\spur}[1]{\not\! #1 \,}
\thispagestyle{empty} \vspace*{1cm} \rightline{BARI-TH/05-517}
\vspace*{2cm}
\begin{center}
  \begin{LARGE}
  \begin{bf}Model independent analysis \\
  \vspace*{0.8cm}
  of a class of $\overline{B}_s^0 $ decay modes
 \vspace*{0.5cm}
  \end{bf}
  \end{LARGE}
\end{center}
\vspace*{8mm}
\begin{center}
\begin{large}
P. Colangelo and R. Ferrandes
\end{large}
\end{center}
\begin{center}
\begin{it}
Istituto Nazionale di Fisica Nucleare, Sezione di Bari, Italy\\
Dipartimento di Fisica,  Universit\'a  di Bari, Italy\\
\end{it}

\end{center}
\begin{quotation}
\vspace*{1.5cm}
\begin{center}
  \begin{bf}
  Abstract\\
  \end{bf}
  \end{center}
\noindent The widths of a class of two-body $\overline{B}_s^0 $
decays  induced by $b \to c \bar u d$ and $b \to c \bar u s$
transitions are determined in a model-independent way, using
$SU(3)_F$  symmetry   and existing information on
   $\bar B \to D_{(s)} P$  and    $\bar B \to D_{(s)} V$ decays, with $P$ and $V$ a light pseudoscalar
   or vector meson. The results are relevant for the $B_s$ physics programmes
 at the hadron  colliders  and at the $e^+ e^-$ factories  running at  the peak of $\Upsilon(5S)$.
\end{quotation}
\newpage
\baselineskip=18pt

In the next few years an intense $B_s$ physics programme will be
pursued at the hadron colliders, the Fermilab Tevatron and  the
CERN LHC, and  at the $e^+ e^-$ factories running at
$\Upsilon(5S)$. The programme includes precise determination of
the $B_s-\overline{B}_s$ mixing parameters and search for CP
violating asymmetries  in $B_s$ decays, with the aim of providing new tests of
 the Standard Model (SM) and   searching for physics beyond SM.  
 The analysis of rare $B_s$ transitions  is another aspect of the research programme, with the same aim of looking for  deviations from SM expectations.

The knowledge of non leptonic $B_s$ decay rates is of prime
importance for working out  the research programme. For example,
$B_s-\overline{B}_s$ mixing can be studied using $B_s$ two-body
hadronic decay modes in addition to semileptonic  modes. It is
noticeable that the widths  of a set of two-body transitions can
be predicted in a model independent way, using the symmetries of
QCD and available information on  $B$ decays. We are referring in
particular to a class of decay modes induced by the quark
transitions $b \to c \bar u d$ and $b \to c \bar u s$, for example
those collected in Table \ref{tab1}.
\begin{table}[hb]
\caption{$SU(3)$ decay amplitudes  for  $\overline{B}_s^0  \to D_{(s)} P$ decays, with $P$
a light pseudoscalar meson.  In the last column the corresponding branching
fractions predicted using the method described in the text are reported.}
    \label{tab1}
    \begin{center}
    \begin{tabular}{| l |c|c|}
  \hline \hline
  decay mode &  amplitude &  BR \\
  \hline
$\overline{B}_s^0 \rightarrow D_s^+\pi^-$ & $V^*_{ud}V_{cb}\,\, T$&  $(2.9\pm0.6)\times 10^{-3}$ \\
$\overline{B}_s^0\rightarrow D^0\overline{K}^0$ & $V^*_{ud}V_{cb}\,\,C$&  $(8.1\pm1.8)\times 10^{-4}$ \\
$\overline{B}_s^0\rightarrow D^0 \eta_8$ &$\frac{1}{\sqrt{6}} \,\, V^*_{us}V_{cb}\,\, (2 C-E)$ &  \\
$\overline{B}_s^0\rightarrow D^0 \eta_0$ &$V^*_{us}V_{cb}\,\, D$ &  \\
$\overline{B}_s^0\rightarrow D^0 \eta$ & &$(2.1\pm1.2) \times 10^{-5}$  \\
$\overline{B}_s^0\rightarrow D^0 \eta^\prime$ & & $(9.8\pm7.6) \times 10^{-6}$\\
$\overline{B}_s^0\rightarrow D^0\pi^0$ & $-\frac{1}{\sqrt{2}}\,\, V^*_{us}V_{cb}\,\, E$&  $(1.0\pm0.3)\times 10^{-6}$ \\
$\overline{B}_s^0\rightarrow D^+\pi^-$ &$V^*_{us}V_{cb}\,\, E$ & $(2.0\pm0.6)\times 10^{-6}$ \\
$\overline{B}_s^0\rightarrow D_s^+ K^-$ &$V^*_{us}V_{cb}\,\, (T+E)$ & $(1.8\pm0.3)\times 10^{-4}$ \\
\hline \hline
    \end{tabular}
   \end{center}
  \end{table}
The key observation is that the various decay modes are governed, in the $SU(3)_F$ limit, by  few independent amplitudes that can be  constrained, both in moduli and in phase differences, from
corresponding $B$ decay processes.

Considering transitions  with  a light pseudoscalar meson
belonging to the octet in the final state, the scheme where the
correspondence can be established involves the three different
topologies
 in $\overline{B}_s^0 $ decays induced by  $b \to c \bar u d (s)$, namely the color allowed topology $T$,
the color suppressed topology $C$ and the $W$-exchange topology
$E$.  The transition in the $SU(3)$ singlet $\eta_0$ involves
another amplitude $D$ in principle not related to the previous
ones. Notice that the identification of the different amplitudes
is not graphical, it is based on $SU(3)$ \cite{Zeppenfeld:1980ex}.
Since $\overline{B}\rightarrow DP$ decays induced by the quark
processes $b\rightarrow c\overline{u}q$ ($q=d$ or $s$) involve a
weak Hamiltonian transforming as a flavor octet, using de Swart's
notation $T^{(\mu)}_\nu$ for the $\nu=(Y,I,I_3)$ component of an
irreducible tensor operator of rank $(\mu)$ \cite{deSwart}, one
can write:
$H_W=V_{cb}V^*_{ud}T^{(8)}_{0\, 1\, -1}+V_{cb}V^*_{us}T^{(8)}_{-1\,\frac{1}{2}\, -\frac{1}{2}}$.
When combined with the initial $\overline{B}$ mesons, which form a
$(3^*)$-representation of SU(3), this leads to $(3^*)$, $(6)$ and
$(15^*)$ representations. These are also the representations
formed by the combination of the final octet light pseudoscalar
meson and triplet D meson. Therefore, using the Wigner-Eckart
theorem for SU(3), the decay amplitudes can be written as linear
combinations of three reduced amplitudes $\langle
\phi^{(\mu)}|T^{(8)}|B ^{(3^*)}\rangle$, with $\mu=3^*,6,15^*$,
which are independent of the quantum numbers $Y,I,I_3$ of the
Hamiltonian and the initial and final states. By appropriate
linear combinations of the  three reduced amplitudes one can
obtain a  correspondence with the three topological diagrams of
the various decay modes. The combinations correspond to  $C$, $T$
and $E$ in Table \ref{tab1},  i.e. the
 color suppressed, color enhanced and W-exchange diagrams, respectively. The $SU(3)$ representation  for  $B$ decays is reported in
Table \ref{tab2}.
\begin{table}[b]
\caption{SU(3) parameterization of  $\bar B \to D_{(s)} P$ decay amplitudes  induced by the $b \to c \bar u d$ and $b \to c \bar u s$ transitions, together with the  experimental results   reported by the Particle Data Group \cite{PDG}.}
    \label{tab2}
    \begin{center}
    \begin{tabular}{|l|c|c|}
  \hline \hline
decay mode &  amplitude & BR \cite{PDG}\\
  \hline
$B^-\rightarrow D^0\pi^-$ & $V^*_{ud}V_{cb}\,\, (C+T)$&  $(4.98\pm0.29)\times10^{-3}$   \\
$\overline{B}^0\rightarrow D^0\pi^0$ & $\frac{1}{\sqrt{2}}V^*_{ud}V_{cb}\,\, (C-E)$& $(2.91\pm0.28)\times10^{-4}$   \\
$\overline{B}^0\rightarrow D^+\pi^-$ & $V^*_{ud}V_{cb}\,\, (T+E)$& $(2.76\pm0.25)\times10^{-3}$   \\
$\overline{B}^0\rightarrow D^+_sK^-$ & $V^*_{ud}V_{cb}\,\,  E$& $(3.8\pm1.3)\times10^{-5}$   \\
$\overline{B}^0\rightarrow D^0 \eta_8$ & $ -\frac{1}{\sqrt{6}} V^*_{ud}V_{cb}\,\, (C+ E)$&    \\
$\overline{B}^0\rightarrow D^0  \eta_0$ & $V^*_{ud}V_{cb}\,\,  D$&    \\
$\overline{B}^0\rightarrow D^0 \eta$ & & $(2.2\pm0.5)\times10^{-4}$    \\
$\overline{B}^0\rightarrow D^0  \eta^\prime$ & &   $(1.7\pm0.4)\times10^{-4}$   \\
 \hline
$B^-\rightarrow D^0K^-$ & $V^*_{us}V_{cb}\,\, (C+T)$ &$(3.7\pm0.6)\times10^{-4}$ \\
  $\overline{B}^0\rightarrow D^0\overline{K}^0$ & $V^*_{us}V_{cb}\,\, C$  & $(5.0\pm1.4)\times10^{-5}$\\
   $\overline{B}^0\rightarrow D^+K^-$ & $V^*_{us}V_{cb} \,\, T$ &$(2.0\pm0.6)\times10^{-4}$ \\
 \hline \hline
    \end{tabular}
   \end{center}
  \end{table}

Considering Table \ref{tab2} one realizes that the three $\bar B \to D K$ experimental rates could
allow to obtain $|T|$, $|C|$ and the phase difference $\delta_C-\delta_T$. This was already observed in
\cite {Xing:2003fe}, and can be recast in the determination of the two independent
isospin amplitudes $A_1$ and $A_0$  for    $I=1$ and $I=0$ isospin $DK$ final states:
${\cal A}(B^- \to D^0 K^-)=\sqrt{2} A_1$,
${\cal A}(\bar B^0 \to D^+ K^-)=\frac{1}{\sqrt 2} \left(A_1 +A_0\right)$
${\cal A}(\bar B^0 \to D^0 \bar K^0)=\frac{1}{\sqrt 2} \left(A_1 -A_0\right)$.
%
\begin{figure}[ht]
\begin{center}
\includegraphics[width=0.40\textwidth] {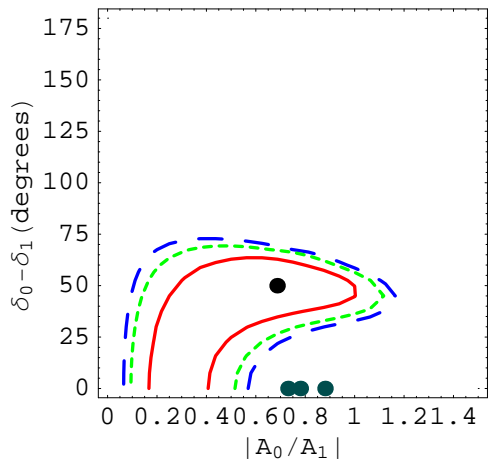}
\vspace*{0mm}
\caption{\baselineskip=15pt
Ratios of isospin amplitudes $A_0/A_1$  for
$\bar B \to DK$ transitions obtained from data in Table  \ref{tab2}.  The contours correspond to the confidence level of $68\%$ (continuous line), $90\%$ (dashed line) and $95\%$ (long-dashed line),
with the dots inside showing the result of the fit. The dots along the $x$ axis correspond to the results of  naive factorization. }
\label{fig:BDK}
\end{center}
\end{figure}
%
Taking into account  the difference
of the $B^-$ and $\bar B^0$ lifetimes:
$\tau_{B^-}=1.671\pm0.018$ ps and $\tau_{B^0}=1.537\pm0.014$ ps,  but  neglecting the  tiny
phase space correction due to the  difference between
 $p_{D^0 K^-}=p_{D^0 \bar K^0}=2280$ MeV and
$p_{D^+ \bar K^-}=2279$ MeV,  with  $p$  the modulus of the three-momentum of one of the
two final mesons in the $B$ rest frame, one would obtain
allowed region for  $A_0/A_1$  at various confidence levels  by minimizing the
$\chi^2$ function for the three branching ratios and plotting the $\chi^2$ contours that correspond
 to a given confidence level,  as done in fig.\ref{fig:BDK}.  Due to the quality of
 the experimental data and  to the   correlation between  $|A_0/A_1|$ and
$\delta_0-\delta_1$, the allowed region is not tightly constrained,  in particular  the phase difference  could be  zero.

We pause here, since we can elaborate once more about
factorization approximations sometimes adopted for computing non
leptonic decays, in this case for $B$ mesons \cite{fact}. In fig.\ref{fig:BDK}
we have shown the predictions by, e.g., naive
factorization, where the decay amplitudes are written in terms of
$K$ and $D$ meson leptonic constants $f_K$ and $f_D$, and the $B
\to D$ and $B \to K$ form factors $F_0$: $ {\cal A}(\bar B^0 \to
D^+ K^-)_F = i \frac{G_F}{\sqrt 2}  V_{cb} V^*_{us} a_1
(m_B^2-m_D^2) f_K F_0^{B \to D}(m_K^2)$ and ${\cal A}(\bar B^0 \to
D^0 \bar K^0)_F = i \frac{G_F}{\sqrt 2} V_{cb} V^*_{us}  a_2
(m_B^2-m_K^2) f_D F_0^{B \to K}(m_D^2)$.  The result of this
approach corresponds to vanishing phase difference; using
 $a_1=c_1+c_2/3$ and $a_2=c_2+c_1/3$, with $c_1$ and $c_2$
the Wilson coefficients appearing in the effective hamiltonian
inducing the decays (for their numerical values we  quote
$a_1=(1.036,1.017,1.025)$ and $a_2=(0.073,0.175,0.140)$ at LO and
at NLO (in NDR and HV renormalization schemes) accuracy,
respectively \cite{buchalla}) we obtain results corresponding  to
the dots along the horizontal axis in fig. \ref{fig:BDK}, which
do not belong to the region permitted by experimental data at
$95\%$ CL. In generalized factorization, where $a_1$ and $a_2$ are
considered as parameters, the phase difference is constrained to
be zero, too. This is allowed by the experimental data on these three channels, but
excluded if one considers all channels, as we shall see below.

\begin{figure}[ht]
\begin{center}
\includegraphics[width=0.40\textwidth] {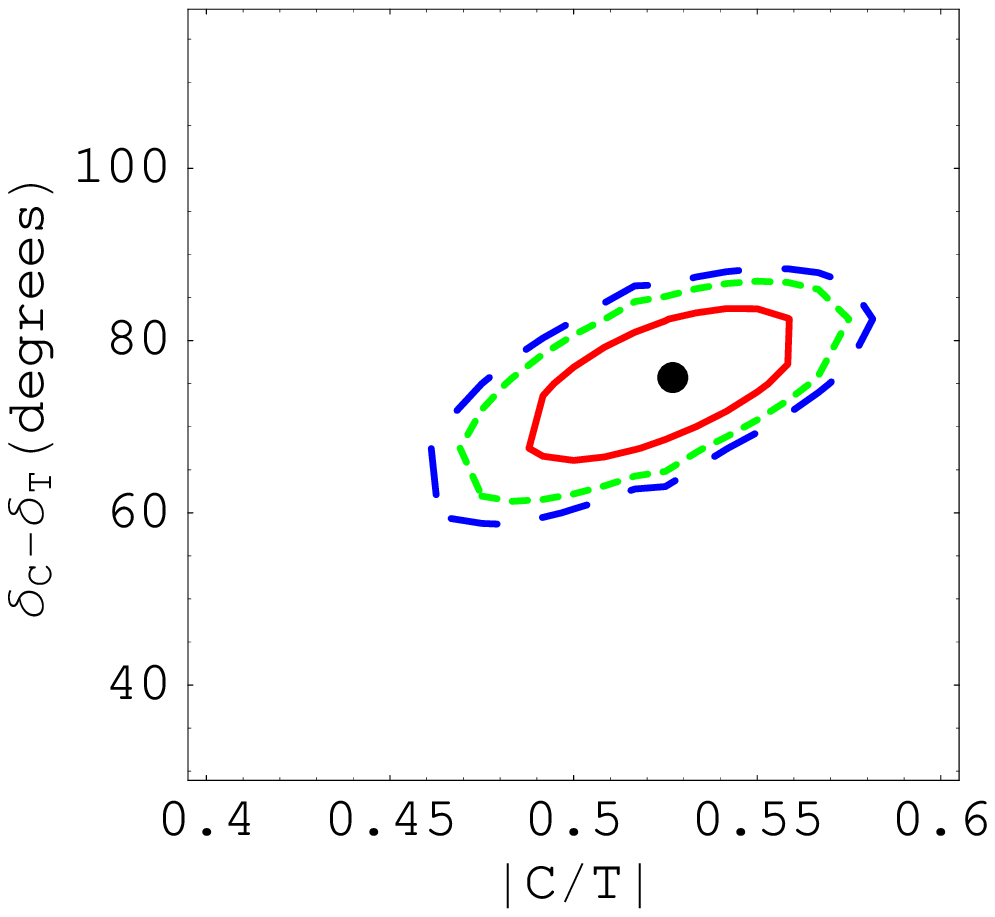} \hspace*{0.3mm}
\includegraphics[width=0.40\textwidth] {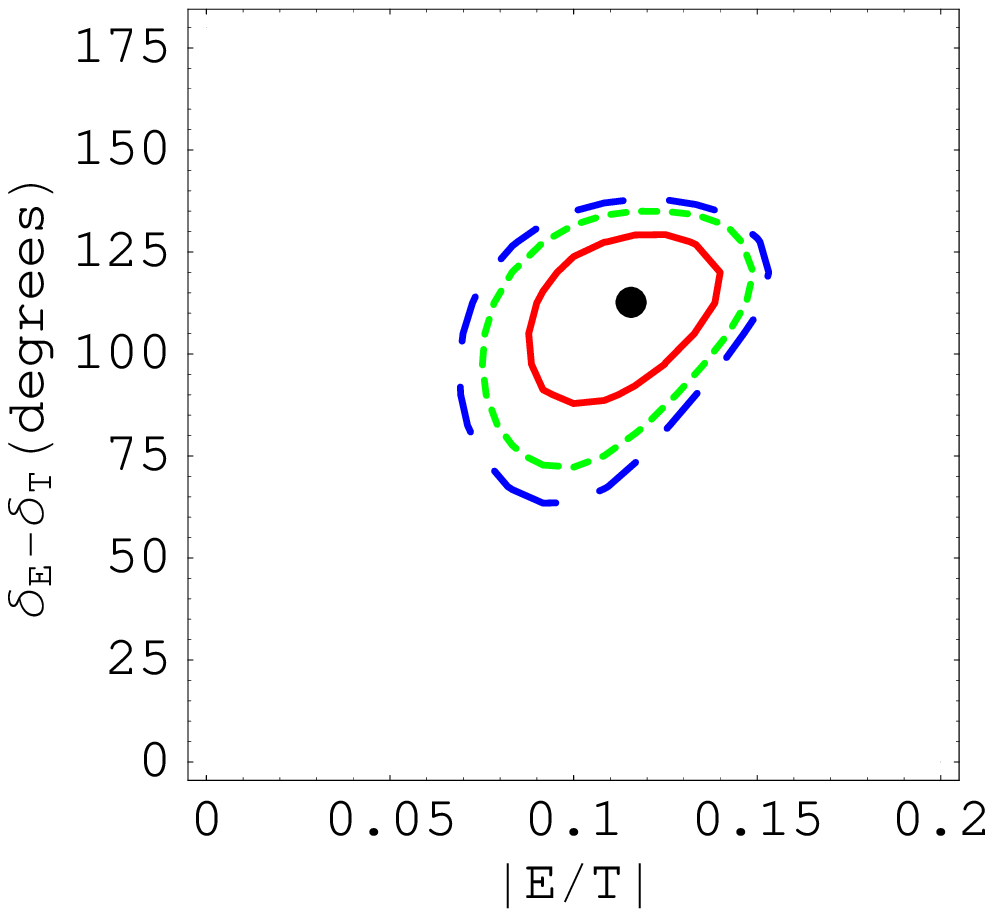}
\vspace*{0mm}
\caption{\baselineskip=15pt
Ratios of SU(3) amplitudes   obtained from data in Table \ref{tab2}.  The contours correspond to the confidence level of $68\%$ (continuous line), $90\%$ (dashed line) and $95\%$ (long-dashed line);
the dots  show the result of the fit.}
\label{fig:BDM}
\end{center}
\end{figure}

Coming  to bounding the decay amplitudes, the four $\bar B \to D \pi$ and $\bar B \to D_s K$ decay rates
cannot determine $C$, $T$, $E$ and their phase differences \cite{neub}.  $\bar B \to D_s K$ only
 fixes the modulus of $E$, which is not small at odds with the expectations by factorization, where
$W$-exchange processes are suppressed by ratios of decay constants
and form factors and are usually considered to be negligible.
Moreover, the presence of $E$ does not allow to directly relate
color  favoured  $T$  or  color  suppressed $C$ decay amplitudes in $D \pi$ and
$D K$ final states. What can be done, however,  is to use all the
information on $\bar B \to D \pi, D_s K$ and $D K$ (7 experimental
data) to determine $T$, $C$ and $E$ (5 parameters).  A similar
attitude has been recently adopted in \cite{kim}. Noticeably, the combined
experimental information is enough accurate to tightly determine the
ranges of variation for all these quantities. In  fig.
\ref{fig:BDM} we have depicted the allowed regions in the $C/T$
and $E/T$ planes, obtained fixing the other variables to their
fitted values, with the corresponding confidence levels. It is worth noticing that
 the phase differences between the various amplitudes
are close to be maximal;  this signals  again  
deviation from naive (or  generalized) factorization, provides contraints to QCD-based approaches
proposed to evaluate non leptonic $B$ decay amplitudes
\cite{Beneke:2000ry, Bauer:2001cu,Keum:2003js}
 and points towards sizeable long-distance effects in $C$ and $E$
 \cite{violations,Chua:2005dt}.
To obtain  the amplitudes
we have fixed  the ratio $|V_{us}/V_{ud}|$  to the experimental
result:  $|V_{us}/V_{ud}|=0.226\pm0.003$,
 and we have taken into account
 the phase space correction due to $p_{D K}$,
 $p_{D\pi}=2306$ MeV and $p_{D_s K^-}=2242$ MeV.
We obtain $|\frac{C}{T}|=0.53 \pm 0.10$,  $|\frac{E}{T}|=0.115\pm0.020$,
$\delta_C-\delta_T=(76\pm12)^\circ$   and   $\delta_E-\delta_T=(112\pm46)^\circ$.
%
%
We have to mention that the accuracy of the fit is not particularly high
since $\chi^2/dof=2.3$,  i.e.  a fit probability  of $10\%$. This is entirely due to a single entry
in Table \ref{tab2},  the branching fraction of $B^- \to D^0 K^-$.  We have reported the PDG value
corresponding to  the average of two measurements, by CLEO  ($(2.92\pm0.84\pm0.28)\times 10^{-4}$) and Belle
($(4.19\pm0.57\pm0.40)\times10^{-4}$)
Collaborations, with the error including a scale factor 1.1. The fit favours a smaller value:
for example, using   the CLEO result  the  $\chi^2/dof$ drops to $0.3$
without sensibly modifying the results.

With the results for the amplitudes  we can determine a number of
$B_s$ decay rates, and   the predictions are collected in Table
\ref{tab1}. The uncertainties in the predicted rates are small; in
particular, the $W$-exchange induced processes   $\overline{B}_s^0
\to D^+ \pi^-, D^0 \pi^0$ are precisely estimated  \cite{aleksan}.

Considering   the decays with  $\eta$ or $\eta'$  in the final state, they
involve the amplitude $D$ corresponding to the transition in a $SU(3)$ singlet $\eta_0$,  and
the $\eta-\eta'$ mixing  angle $\theta$ (in a one angle mixing scheme):
\bea\label{eq1}
A(\overline{B}^0 \rightarrow D^0\eta)
&=&\cos\theta[-\frac{1}{\sqrt{6}}(C+E)]-\sin\theta\frac{D}{\sqrt{3}}\nn\\
A(\overline{B}^0 \rightarrow D^0\eta')
&=&\sin\theta[-\frac{1}{\sqrt{6}}(C+E)]+\cos\theta\frac{D}{\sqrt{3}}
\,\,\,. \eea If we use the value   $\theta=-15.4^0$ for the mixing
angle \cite{feldmann}, we obtain $|\frac{D}{T}|=0.41\pm0.11$
without sensibly constraining the $D-T$ phase difference,
$\delta_D-\delta_T=-(25 \pm 51)^\circ$. Corresponding
$\overline{B}_s^0 $ decay rates are predicted consequently.

\begin{figure}[ht]
\begin{center}
\includegraphics[width=0.40\textwidth] {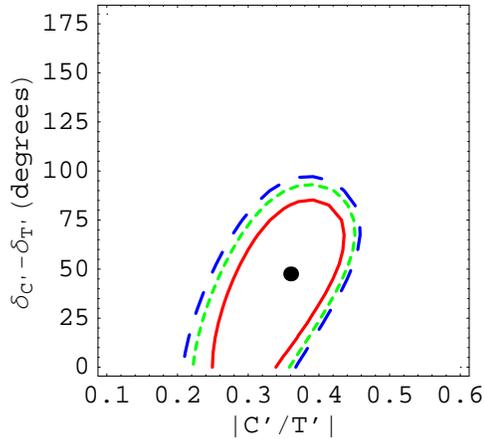}
\vspace*{0mm}
\caption{\baselineskip=15pt
Ratio of $C^\prime$ and $T^\prime$ amplitudes for
$\bar B \to DV$ transitions.  The contours correspond to the same confidence level  as in fig. \ref{fig:BDM}; the dot shows the result of the fit. }
\label{fig:BDV}
\end{center}
\end{figure}

 The key of the success of the  programme of predicting $B_s$ decay rates is  the
 small number of amplitudes in comparison  to the available data, a feature which is  not common
 to all processes.  Considering  $b \to c \bar u d (s)$ induced  transitions,  one could look at
 the case of one light vector meson in the final state,  with  the same $SU(3)$ decomposition
 reported in Tables \ref{tab1}, \ref{tab2} (we denote by a prime the amplitudes involved in this case).
$B$ decay data are collected in Table \ref{tab3}.  The difference with respect to the previous case is that
the W-exchange mode $\bar B^0 \to D^+_sK^{*-}$  has not been observed, yet, therefore the $E^\prime$
amplitude is poorly determined considering only  the other modes.
  Taking into account phase space corrections due to
$p_{D \rho}=2235$ MeV and $p_{D K^*}=2211$ MeV, we obtain
$|\frac{C'}{T'}|=0.36 \pm 0.10$, $|\frac{E'}{T'}|=0.29 \pm 0.37$,
$\delta_{C'}-\delta_{T'}= (48\pm67)^\circ$ and
$\delta_{E'}-\delta_{T'}=(96\pm56)^\circ$, with a fit probability
of $85\%$. The allowed region in the $C^\prime/T^\prime$ plane, fixing all the other variables to
their fitted values,  is depicted in fig.\ref{fig:BDV}. In this case the phase difference
$\delta_{C^\prime}-\delta_{T^\prime}$ can vanish.
%
%
%
\begin{table}[ht]
\caption{Experimental results for the branching fractions of $\bar
B \to D_{(s)}  V$  decays induced by the $b \to c \bar u d$ and $b \to c
\bar u s$ transitions  as reported by the Particle Data Group \cite{PDG}. The  predictions for
$\overline{B}_s^0 $ decays obtained using the method described in the text are reported in the last
column.}
    \label{tab3}
    \begin{center}
    \begin{tabular}{|l|c| |l|c|}
  \hline \hline
decay mode &   BR \cite{PDG} & decay mode & BR\\
\hline
$B^-\rightarrow D^0\rho^-$ &  $(1.34\pm0.18)\times10^{-2}$&$\overline{B}_s^0  \rightarrow D _s^+ \rho^{-}$ &   $(7.2\pm 3.5)\times10^{-3}$ \\
$\overline{B}^0\rightarrow D^0\rho^0$ &  $(2.9\pm1.1)\times10^{-4}$&$\overline{B}_s^0  \rightarrow D^0\overline{K}^{*0}$ & $(9.6\pm2.4)\times10^{-4}$ \\
$\overline{B}^0\rightarrow D^+\rho^-$    &  $(7.7\pm1.3)\times10^{-3}$& &  \\
$\overline{B}^0\rightarrow D_s K^{*-}$    &  $<  \,9.9 \times10^{-4}$& & \\
 \hline
 $B^-\rightarrow D^0 K^{*-}$ &  $(6.1\pm2.3)\times10^{-4}$&$\overline{B}_s^0 \rightarrow D^0 \rho^{0}$ &   $(0.28\pm 1.4)\times10^{-4}$ \\
$\overline{B}^0\rightarrow D^0 \bar K^{*0}$ &  $(4.8\pm1.2)\times10^{-5}$&$\overline{B}_s^0  \rightarrow D^+ \rho^-$ & $(0.57\pm 2.8)\times10^{-4}$ \\
$\overline{B}^0\rightarrow D^+ K^{*-}$    &
$(3.7\pm1.8)\times10^{-4}$&
$\overline{B}_s^0 \rightarrow D_s^+K^{*-}$ &  $(4.5\pm3.1)\times10^{-4}$ \\
\hline \hline
      \end{tabular}
   \end{center}
  \end{table}
The predictions for $\overline{B}_s^0 $ decay rates are collected
in Table \ref{tab3}: as anticipated, the accuracy is not high for
$W-$exchange induced decays.  On the other hand,  the prediction
for the rate of $\overline{B}^0 \to D_s K^{*-}$:  ${\cal
B}(\overline{B}^0 \to D_s K^{*-})=(1\pm5)\times 10^{-4}$, is
compatible with the upper bound in Table \ref{tab3}.

Considering other decay modes induced by the same quark
transitions, namely  $\bar B \to D^*_{(s)}ÊP$ and $\bar B \to
D^*_{(s)}ÊV$ decays, the  present experimental data are not
precise enough to sensibly constrain the independent amplitudes
and to provide stringent predictions for $B_s$.  As soon as the
experimental accuracy will improve, a similar analysis  will be
possible to describe  $\overline{B}_s^0  \to D^*_{(s)}ÊP$ modes,
while the three helicity
 $\bar B \to D^*_{(s)}ÊV$ amplitudes will be needed to
 determine the corresponding  $B_s$ decays.

Let us finally comment on the possible role of $SU(3)_F$ breaking
terms  that can modify our predictions. Those effects are not
universal,  and in
general cannot be reduced to well defined and predictable
patterns without new assumptions. Their
parametrization  would introduce additional  quantities \cite{Gronau:1995hm}  that  at present
cannot be sensibly bounded since  their effects seem to be
smaller than  the experimental uncertainties.
Therefore they can be  neglected until the experimental
errors remain at the present level.  It will be interesting to
investigate their role when the $B_s$ decay rates will be measured and
more precise $B$ branching fractions  will be available.

\vspace*{1cm}
\noindent {\bf Acnowledgments}\\
\noindent
We thank F. De Fazio for discussions. We acknowledge partial support from the EC Contract No.
HPRN-CT-2002-00311 (EURIDICE).

\end{document}